\documentstyle[preprint,aps]{revtex}

\def\head{\vspace{-7cm}\begin{flushright} DESY 95-131\\
                                          July 4, 1995\\[1cm]
                       \end{flushright}}

\def\tmtwo{$\times10^{-2}$}
\def\tmthree{$\times10^{-3}$}
\def\tmfour{$\times10^{-4}$}
\def\tmfive{$\times10^{-5}$}
\def\tmsix{$\times10^{-6}$}
\def\tmseven{$\times10^{-7}$}
\def\tmeight{$\times10^{-8}$}

\newcommand{\ba}{\begin{array}}
\newcommand{\ea}{\end{array}}
\newcommand{\bd}{\begin{displaymath}}
\newcommand{\ed}{\end{displaymath}}
\newcommand{\be}{\begin{equation}}
\newcommand{\ee}{\end{equation}}
\newcommand{\bea}{\begin{eqnarray}}
\newcommand{\eea}{\end{eqnarray}}

\def\qb{\bar{q}}

\def\cb{\bar{c}}
\def\sb{\bar{s}}

\def\to{\rightarrow}       

\def\ln{\mbox{$\ell n$}}   
\def\sin{\mbox{ sin}}
\def\cos{\mbox{ cos}}
\def\Sum{\displaystyle\sum}
\def\eff{{\rm eff}}
\def\Heff{{\cal H}_{\rm eff}}

\def\me{{\bf m}}
\def\r{{\bf r}}

\def\eq{}

\def\bra{\langle}
\def\ket{\rangle}

\def\ae{\alpha_{em}}
\def\a{\alpha}
\def\b{\beta}
\def\g{\gamma}
\def\d{\delta}

\begin{document}
\head
\begin{center} \begin{Large} \begin{bf}
Electroweak and Strong Penguins in $\bf {B^{\pm, 0}\rightarrow \pi\pi, \pi K}$
and $\bf KK$ Decays

\end{bf} \end{Large} \end{center}
\vspace{1cm}
\begin{center}
    G.\ Kramer$^a$,
    W.\ F.\ Palmer$^b$\footnote{Supported in part by the US
            Department of Energy under contract DOE/ER/01545-605.},\\
      \vspace{0.3cm}
        $^a$II. Institut f\"ur Theoretische Physik\footnote{Supported by
            Bundesministerium f\"ur Forschung und Technologie,
            05\,6\,HH\,93P(5), Bonn, Germany and EEC Program
            ``Human Capital and Mobility" Network ``Physics at High Energy
            Colliders" CHRX-CT93-0357 (DG 12 COMA)}\\
            der Universit\"at Hamburg,\\
        D--22761 Hamburg, Germany\\

  \vspace{0.3cm}
        $^b$Department of Physics, The Ohio State University, \\
        Columbus, Ohio 43210, USA\\
        \end{center}
  \vspace{1cm}
\eject
\noindent {\bf Abstract}\\
\parbox[t]{\textwidth}{
We calculate CP-violating rates and asymmetry parameters in charged and neutral
$B\to \pi\pi, \pi K$ and $\bar K K$ decays arising from the interference of
tree and penguin (strong and electroweak) amplitudes with different strong and
CKM phases. The perturbative strong (electroweak) phases develop at order
$\alpha_s$ ($\alpha_{em}$) from absorptive parts of one-loop matrix elements of
the next-to-leading (leading) logarithm corrected effective Hamiltonian. The
BSW model is used to estimate the hadronic matrix elements.  Based on this
model, we find that the effect of strong phases and penguins is substantial in
most channels, drastic in many.  However, a measurement of the time dependence
parameter $a_{\epsilon+\epsilon'}$ in the $\pi^+\pi^-$ channel is only
influenced at the 20\% level by the complication of the penguins.    Recent
flavor  sum rules developed for $B^{0,\pm} \rightarrow \pi\pi, \pi K, K \bar K$
amplitudes are tested in this model. Some are well satisfied,  others badly
violated, when  electroweak penguins are included.}
\newpage

\section{Introduction}

So far CP violation \cite{Jarl} has been detected only in processes related to
$K^0 - \bar K^0$ mixing \cite{CCFT} but considerable efforts are being made to
find it in $B$ decays.  Partial rate and time dependence
asymmetries are the leading signals if the CKM \cite{CKM} model of CP
violation is the
correct guide.  The next generation of $B$ experiments will then make detailed
probes of rates and time dependence from which, in principle, a definitive
 test of
the CKM scheme can be made in a model independent way. This
amounts to testing whether the unitarity triangle closes, that is, whether
three generations and a single phase suffice to describe CP violation in the K
and B systems.

CP violation in the CKM model is either `direct,' in which two amplitudes for
the same process have different weak and strong phases, or `indirect,' in which
one of the two interfering amplitudes proceeds through the mixing of the
neutral
$B$ and $\bar B$ mesons.  Rate asymmetries between charge conjugate $B^\pm$
exclusive channels are purely
direct. Their advantage is that they do not require complicated time dependence
measurements, involving tagging, and they are definitive signals of CP
violation
if seen.  Their disadvantage is that a model is needed to
calculate the strong phase if CKM phase information is to be extracted.
Typically these asymmetries arise when tree and penguin amplitudes interfere.
A further complication is that direct and indirect CP violation occur
simultaneously in neutral decays when there are penguin as well as tree
amplitudes in addition to mixing.  A recent treatment of this complication
can be found in reference \cite{WP}.

Various authors have shown how to extract CKM phases from certain sets of
measurements in a fairly model independent way.  Gronau and London \cite{GL}
have shown that measurements of all charge states in $B\rightarrow \pi\pi$, as
well as the time dependence  of $B\rightarrow \pi^+\pi^-$ are required to
obtain the CKM phases. Nir and Quinn \cite{NQ} have extended this analysis to
the $K \pi$ system. Finally, Gronau et al. \cite{GRL} have argued that by
using SU(3) relations, there is enough information from rates alone to
determine the CKM phases without measurements of time dependence. Unfortunately
this analysis relied heavily on the assumption that electroweak penguins were
negligible.  Subsequently Deshpande and He \cite{k6,k7} showed that this
analysis which relied on SU(3) symmetry and the particular structure of the
strong penguins ceases to be true when electroweak penguins are included.
Recently the authors of reference \cite{GRL} have modified and extended their
approach to include electroweak (EW) penguins and $B_s$
transitions \cite{GRLpen}.

In practice a full determination of all rates and time dependence is a
formidable experimental task.  Thus in the first stages input from models will
be useful to guide the analysis to the measurements which are most easily made
and which can give the best early answers.  This requires a consistent
modelling of tree and penguin amplitudes including strong phases.  The strong
phase is generated by final state interactions. At the quark level the strong
interaction effects can be modeled perturbatively, following Bander, Silverman
and Soni \cite{BSS}, by the absorptive part of penguin diagrams.

Recently we have developed a model for the strong phases of the penguins and
applied it to rate asymmetries in charged $B\rightarrow PP, PV, VV$
channels \cite{k9,k10}. In order to systematically take into account the
$O(\alpha_s)$ penguin matrix elements, we base our treatment on the
next-to-leading logarithmic short distance corrections evaluated by Buras et
al. \cite{k2}. In this work we have also included the $O(\alpha_{em})$
electroweak penguins in order to investigate their effect on rates,
asymmetries, and the SU(3) sum rules.

Having modeled the tree and penguin operators, we use factorization and the BSW
current matrix elements to calculate rates and asymmetries.    In this note we
will apply this model to neutral and charged $\pi\pi, \pi K, KK$ channels. Of
course, in order to make definitive predictions we need the CKM parameters as
input.  A recent study of the experimental data constraining these parameters
is the work of Ali and London \cite{Ali} which shows a preferred solution of
$\rho=-0.12, \eta=0.34$ in the Wolfenstein representation. We use this solution
as an illustrative example of how significant strong and electroweak penguins
can be in this model. Of course other still possible values for $\rho$ and
$\eta$ will lead to somewhat different predictions.

The remainder of this paper is organized as follows. In Sect.~2 we describe the
effective weak Hamiltonian and the evaluation of the hadronic matrix elements.
In Sect.~3 we give a short account of ref. \cite{WP} needed for our
calculation.
The final results for the branching ratios and rate differences are discussed
in Sect.~4. In this section we also analyse the SU(3) sum rules.
In Sect.~5 we end with a short summary and draw some
conclusions.

\section{The Effective Hamiltonian}
In the next two subsections we present the short distance Hamiltonian and
the quark-level matrix elements. In subsection 2.3 we describe the evaluation
 of the
hadronic matrix elements which are relevant for the $PP$ final states.

\subsection{Short Distance QCD Corrections}
For calculations of CP-violating observables it is most convenient to
exploit the unitarity of the CKM matrix and split the effective weak
Hamiltonian into two pieces, one proportional to
$v_u\equiv V_{ub}V_{us}^\ast$ (or $V_{ub}V_{ud}^\ast$ in the case of
$b\to d$ transitions) and the other one proportional to
$v_c\equiv V_{cb}V_{cs}^\ast$ (or $V_{cb}V_{cd}^\ast$ correspondingly),
\be
\Heff = 4 \frac{G_F}{\sqrt{2}} \left( v_u \Heff^{(u)} + v_c \Heff^{(c)} \right)
\ . \label{def_of_Hq} \ee
The two terms ($q = u,c$)
\be \Heff^{(q)} = \sum_i c_i(\mu) \cdot O_i^{(q)}\ , \ee
differ only by the quark content of the local operators,
and for our purposes it is sufficient to consider only
the following four-quark operators:
\be\ba{llllll}
O_1^{(q)} & = & \sb_\a \g^\mu L q_\b \cdot \qb_\b \g_\mu L b_\a \ , \ \ &
O_2^{(q)} & = & \sb_\a \g^\mu L q_\a \cdot \qb_\b \g_\mu L b_\b \ , \\ \
O_3 & = & \sb_\a \g^\mu L b_\a \cdot \Sum_{q'}\qb_\b'\g_\mu L q_\b'\ , \ \  &
O_4 & = & \sb_\a \g^\mu L b_\b \cdot \Sum_{q'}\qb_\b'\g_\mu L q_\a'\ , \\
O_5 & = & \sb_\a \g^\mu L b_\a \cdot \Sum_{q'}\qb_\b'\g_\mu R q_\b'\ , \ \  &
O_6 & = & \sb_\a \g^\mu L b_\b \cdot \Sum_{q'}\qb_\b'\g_\mu R q_\a'\ , \\
O_7 & = & \sb_\a \g^\mu L b_\a \cdot \Sum_{q'}e_{q'}\qb_\b'\g_\mu R q_\b'\ , \
 \  &
O_8 & = & \sb_\a \g^\mu L b_\b \cdot \Sum_{q'}e_{q'}\qb_\b'\g_\mu R q_\a'\ , \\
O_9 & = & \sb_\a \g^\mu L b_\a \cdot \Sum_{q'}e_{q'}\qb_\b'\g_\mu L q_\b'\ , \
 \  &
O_{10} & = & \sb_\a \g^\mu L b_\b \cdot \Sum_{q'}e_{q'}\qb_\b'\g_\mu L q_\a'\ .
\label{operators}
\ea \ee
where $L$ and $R$ are the left- and right-handed projection operators.
The $q'$ run over the quark operators that are active at the scale $\mu=O(m_b),
(q'\epsilon\{u,d,s,c,b\})$.  The operators $O_3,\ldots,O_6$ arise from (QCD)
penguin diagrams which contribute at order $\a_s$ to the initial values of the
coefficients at $\mu \approx M_W$\cite{k2} or through operator mixing
during the renormalization group summation of short distance QCD corrections
\cite{QCD_SD}. The usual tree-level $W$-exchange corresponds to $O_2$ (with
$c_2(M_W) = 1+O(\a_s)$). Similarly, $O_7,\ldots,O_{10}$ arise from electroweak
penguin diagrams.

The renormalization group evolution from $\mu \approx M_W$ to $\mu
\approx m_b$ has been evaluated
in leading logarithmic (LL) order in the electromagnetic coupling
and in
next-to-leading logarithmic (NLL) precision in the strong coupling $\alpha_s$.
The
full 10x10
matrices for the anomalous dimensions $\gamma_i$ are given by Buras et al.
in references
\cite{k3,k4,k5}.  The Wilson coefficients $c_i$ obtained depend on the
renormalization scheme used.  This renormalization scheme dependence can be
isolated in terms of the matrices $\r^{s,ew}_{ij}$ by writing \cite{k2,k5,k1}

\be c_j(\mu) = \sum_i \bar{c}_i(\mu)  \left[
\d_{ij} - \frac{\a_s(\mu)}{4\pi} \r^s_{ij}
- \frac{\ae(\mu)}{4\pi} \r^{ew}_{ij}\right] \ , \ee
 where the coefficients
$\cb_j$ are scheme independent at this order. The matrix elements
$\r^{s,ew}_{ij}$ have been evaluated in  references \cite{k1,k2,k5,k6}.
To obtain numerical values for the $\bar c_i$ we must specify the input. We
choose $\alpha_s(M_z)=0.118$, $\alpha_{em}(M_z)=1/128$ and
$\mu=m_b=\,4.8~GeV$. Then we have \cite{k2,k7}

\be\ba{llllll}
\cb_1 & = & - 0.324\ , \ \ & \cb_2 & = & 1.15\ , \\
\cb_3 & = & 0.017\ , \ \ & \cb_4 & = & -0.038\ , \\
\cb_5 & = & 0.011\ , \ \ & \cb_6 & = & -0.047\ , \\
\cb_7 & = & -1.05$\tmfive$\ , \ \ & \cb_8 & = & -3.84$\tmfour$\ , \\
\cb_9 & = & -0.0101\ , \ \ & \cb_{10} & = & 1.96$\tmthree$\ , \\
\ea\ee

Other values using slightly different input can be found in \cite{k1,k2}.
The electroweak coefficient $\cb_9$, as noted in reference \cite{k7}, is
not much smaller than the strong penguins; its major contribution arises
from the $Z$ penguin.

\subsection{Quark-level Matrix Elements}
Working consistently at NLL precision, the matrix elements of $\Heff$ are
to be treated at the one-loop level in order to cancel the scheme
dependence from the renormalization group evolution. The one-loop matrix
elements can be rewritten in terms of the tree-level matrix elements of
the effective operators \cite{k1,k8,k9,k10}
\be \bra sq'\qb'\vert \Heff^{(q)}\vert b\ket =
\sum_{i,j} c_i(\mu)
\left[ \d_{ij} + \frac{\a_s(\mu)}{4\pi} \me_{ij}^s(\mu,\ldots)
+\frac{\ae(\mu)}{4\pi} \me_{ij}^{ew}(\mu,\ldots)\right]
\bra sq'\qb'\vert O_j^{(q)}\vert b\ket^{\rm tree} \ . \label{me1}\ee
The functions $\me^{s,ew}_{ij}$ are determined by the corresponding
renormalized
one-loop diagrams and depend in general on the scale $\mu$, on the quark
masses and momenta, {\it and} on the renormalization scheme. The various
one-loop diagrams can be grouped into two classes: {\it vertex-corrections},
where a gluon connects two of the outgoing quark lines,
and {\it penguin} diagrams, where a quark-antiquark line closes a loop
and emits a gluon, which itself decays finally into a quark-antiquark pair.
The full matrices $\me^{s,ew}_{ij}$ have been worked out by Fleischer
 \cite{k1,k8} and Deshpande and He \cite{k6}.

When expressing the rhs of \eq(\ref{me1}) in terms of the renormalization
scheme independent coefficients $\cb_i$, the effective coefficients multiplying
the matrix elements $\bra sq'\qb'\vert O_j^{(q)}\vert b\ket^{\rm tree}$
become
\be c_j^\eff \equiv \cb_j +  \sum_i \cb_i \cdot
[\frac{\a_s}{4\pi}\left( \me^s_{ij}-\r^s_{ij} \right)
+\frac{\ae}{4\pi}\left( \me^{ew}_{ij}-\r^{ew}_{ij} \right)] \ . \ee
The renormalization scheme dependence, which is present in $\me_{ij}$
and $\r_{ij}$, explicitly cancels in the combinations
 $\me^{s,ew}_{ij}-\r^{s,ew}_{ij}$
\cite{k2,k1,k8}.

The effective coefficients multiplying
the matrix elements $\bra sq'\qb'\vert O_j^{(q)}\vert b\ket^{\rm tree}$
become

\bea
c_3^\eff & = & \cb_3 - \frac{1}{N} \frac{\a_s}{8\pi} (c_t + c_p)
      \nonumber\\
c_4^\eff & = & \cb_4 +  \frac{\a_s}{8\pi} (c_t + c_p)
      \nonumber\\
c_5^\eff & = & \cb_5 - \frac{1}{N} \frac{\a_s}{8\pi} (c_t + c_p)
      \nonumber\\
c_6^\eff & = & \cb_6 +  \frac{\a_s}{8\pi} (c_t + c_p)
      \nonumber \\
c_7^\eff & = & \cb_7 +  \frac{\ae}{8\pi} c_e
      \nonumber\\
c_8^\eff & = & \cb_8
      \nonumber\\
c_9^\eff & = & \cb_9 +  \frac{\ae}{8\pi} c_e
      \nonumber\\
c_{10}^\eff & = & \cb_{10}
     , \label{c3456} \eea
where we have separated the contributions $c_t$ and $c_p$ from the ``tree''
operators $O_{1,2}$ and from the strong penguin operators $O_{3\cdots6}$,
respectively. $c_e$, given below, comes from the electroweak penguins.

In addition to the contributions from penguin diagrams with insertions of
the tree operators $O^{(q)}_{1,2}$
\be
c_t = \cb_2 \cdot \left[\frac{10}{9}+\frac{2}{3} \ln \frac{m_q^2}{\mu^2}
- \Delta F_1 \Bigl(\frac{k^2}{m_q^2}\Bigr) \right]\ , \label{ct}
 \ee
where $\Delta F_1$ is defined in \cite{k9}, we have evaluated
the penguin diagrams for the matrix elements of the penguin operators
\cite{k9}:
\bea
c_p & = & \cb_3 \cdot \left[
       \frac{280}{9}
       + \frac{2}{3} \ln \frac{m_s^2}{\mu^2}
       + \frac{2}{3} \ln \frac{m_b^2}{\mu^2}
       - \Delta F_1\Bigl(\frac{k^2}{m_s^2}\Bigr)
       - \Delta F_1\Bigl(\frac{k^2}{m_b^2}\Bigr) \right] \nonumber \\
& + & (\cb_4+\cb_6)\cdot \sum_{j=u,d,s,\ldots} \left[
       \frac{10}{9}+\frac{2}{3} \ln \frac{m_j^2}{\mu^2}
      - \Delta F_1\Bigl(\frac{k^2}{m_j^2}\Bigr) \right] \ ,
\eea

For the electroweak penguins we consider only those arising from the
insertion of the tree operators.  The corresponding coefficient is
given by

\bea
c_e & = & \frac{8}{9}(3\cb_1+\cb_2)(\frac{10}{9} + \frac{2}{3} \ln
        \frac{m_q^2}{\mu^2}-\Delta F_1(\frac{k^2}{m_q^2}) ,
\eea

Note that the coefficients $c_i^\eff$ depend on $k^2$ and, as we shall
see later, on the $q\bar{q}$ states that are included in the sum over
the intermediate states.

\subsection{Hadronic Matrix Elements in the BSW Model}
To take into account long distance QCD effects which build up the
hadronic final states, we follow Bauer, Stech and Wirbel \cite{BSW}:
With the help of the factorization hypothesis the three-hadron
matrix elements are split into vacuum-meson and meson-meson matrix elements
of the quark currents entering in $O_1,\ldots,O_{10}$. In addition, OZI
suppressed form factors and annihilation terms are neglected.
In the BSW model, the meson-meson matrix elements of the currents are
evaluated by overlap integrals of the corresponding wave functions and
the dependence on the momentum transfer (which is equal to the mass of
the factorized meson) is modeled by a single-pole ansatz.
As a first approximation, this  calculational scheme
provides a reasonable method for estimating the relative size and
phase of the tree and penguin terms that give rise to the CP-violating
signals.

When pseudoscalars are involved there are additional contributions
from the $(V+A)$
penguin operator $O_6$ and $O_8$: After Fierz reordering and factorization
they contribute terms which involve a matrix element of the
quark-density operators between a pseudoscalar meson and the
vacuum. For $O_6$, for example, this is given by
\be
\bra P_1 P_2\vert O_6 \vert B \ket = - 2 \Sum_q \Bigl(
\bra P_1 \vert \bar q b_L \vert 0\ket \bra P_2 \vert\sb q_R \vert B \ket +
\bra P_2 \vert \bar q b_L \vert 0\ket \bra P_1 \vert\sb q_R \vert B \ket
\Bigr) \ . \ee

Using the Dirac equation, the matrix elements entering here can be
rewritten in terms of those involving usual $(V\!-\!A)$ currents,
\be
   \bra P_1 P_2 \vert O_6 \vert B \ket =
   R[P_1,P_2] \bra P_1 P_2 \vert O_4 \vert B \ket \ ,
\label{O6_vs_O4} \ee
with
\be
R[P_1,P_2] \equiv
\frac{2M^2_{P_1}}{(m_{q1}+m_{\bar q1})(m_b - m_{q2})} \ .
\label{def_of_R} \ee
Here, $m_{q1}$ ($m_{\bar q1}$) and $m_{q2}$ are the current masses of the
(anti-)quark in the mesons $P_1$ and $P_2$, respectively.
We use the quark masses $m_u=m_d= 10$~MeV, $m_s=200$~MeV, $m_c = 1.5$~GeV
and $m_b = 4.8$~GeV.  The same relations work for $O_8$.

Finally, one arrives at the form
\bea \bra P_1\,P_2\vert \Heff^{(q)}\vert B\ket
& = & Z^{(q)}_1 \bra P_1 \vert j^{\mu}\vert 0 \ket
          \bra P_2 \vert j_{\mu} \vert B\ket \nonumber \\
& + & Z^{(q)}_2 \bra P_2 \vert j'^{\mu}\vert 0 \ket
          \bra P_1 \vert j'_{\mu} \vert B\ket \ ,
\label{def_of_Z12} \eea
where $j_\mu$ and $j'_\mu$ are the corresponding (neutral or charged)
$V\!-\!A$ currents. The factorization coefficients $Z^{(q)}_1$ and
$Z^{(q)}_2$ are listed in the appendix.
In terms of the form factors $F_{0}$ for the current matrix
elements defined by BSW \cite{BSW}, this yields
\bea \bra P_1 P_2 \vert \Heff^{(q)}\vert B\ket  & = &
        Z^{(q)}_1 (M_B^2-M_2^2) f_{P_1} F_0(M_1^2)
      + Z^{(q)}_2 (M_B^2-M_1^2) f_{P_2} F_0(M_2^2) \
   \eea
$M_B$ is the mass of the decaying $B$ meson and $f_P$ is the
decay constant of the pseudoscalar mesons in the final state.

Concerning how $1/N$ terms are treated in the coefficients (see
\eq(\ref{c3456}) and \eq(\ref{def_of_ai})), it is well known \cite{Stone} that
this model has problems accounting for the decays with branching ratios which
are proportional to the combination $ \bar{c}_1 + \bar{c}_2/N$. This is due to
the rather small absolute value of this particular combination when using the
short-distance QCD corrected coefficients. An analogous effect is also known in
nonleptonic D decays \cite{BSW}, and several authors advocated a modified
procedure to evaluate the factorized amplitudes \cite{BSW,Buras2}: There, only
terms which are dominant in the $1/N$ expansion are taken into account.
Recently there has been much discussion in the literature concerning these
issues.  Fits to measured branching ratios indicate that the effective $a_1$
coefficient has the opposite sign from that expected if the $1/N$ terms are
completely cancelled by non-factorizable terms, as if the cancellation were
incomplete, an effect mimicked by taking $N = 2$.  In this work we shall quote
results for $N=\infty$ and $N = 2$.

The strong phase shifts are generated in our model only by the
absorptive parts (hard final state interactions) of the quark-level
matrix elements of the effective Hamiltonian. Of course, when factorizing
the hadronic matrix elements, all information on the crucial value of the
momentum transfer $k^2$ of the gluon in the penguin diagram is
lost. While there has been an attempt \cite{SW}
to model a more realistic momentum distribution by taking into account
the exchange of a hard gluon, we will use here for simplicity only a
fixed value of $k^2$. From simple two body kinematics \cite{Deshpande}
or from the investigations in ref.~\cite{SW} one expects $k^2$ to be
typically in the range
\be
\frac{m_b^2}{4}\stackrel{<}{\sim} k^2 \stackrel{<}{\sim}\frac{m_b^2}{2}\ .
\label{kk_range} \ee

The results we shall present are sensitive to $k^2$ in this range
because the $c\bar{c}$ threshold lies between these limits. Arguments
have been made that the lower limit is a more appropriate choice
\cite{SEW}. In this work we follow \cite{k9} and choose the upper
limit for our numerical presentation in the tables.

\def\etau{\eta_u}
\def\etas{\eta_s}
The factorization coefficients $Z^{(q)}_{1,2}$ defined in
\eq(\ref{def_of_Z12}) are listed in Tab. 1 a,b for
the $B\to PP$ channels of interest.  Here we add tree and penguin
contributions.  It is understood that they have different CKM factors
which must be inserted.
Colour suppressed terms may readily be included in the coefficients
\be
a_i \equiv c_i^\eff + \frac{1}{N} c_j^\eff
\label{def_of_ai} \ee
where $\{i,j\}$ is any of the pairs $\{1,2\}$, $\{3,4\}$, $\{5,6\}$,
$\{7,8\}$,or $\{9,10\}$.
In Tab. 1 a,b we have adopted the convention of including
factors of $\sqrt{2}$ associated with a neutral meson $P_2$. They arise
either from current matrix elements between $P_2$ and $B$ (Tab. 1 a),
or from the definition of the decay constants for $P_2$ (Tab. 1 b).
Care should be taken with the latter since these factors are sometimes
absorbed into the decay constants (e.g. as tabulated in \cite{BSW}).

Our phase convention for the mesons states relative to the weak
Hamiltonian is defined by:
\be
 [\pi^+, \pi^0, \pi^-] = [\bar u d, \frac{d \bar d - u \bar u}{\sqrt{2}},
 d \bar u]
\ee

\be
(K^+,K^0) = (u \bar s,~~ d \bar s),~~~(K^-, K^0) = (s \bar d, ~~s \bar u)
\ee

\be
(B^-, B^0) = (b \bar u,~~ b \bar d),~~~(B^+, B^0) = (\bar b u,~~ \bar b d)
\ee
\section{CP-violating Observables}

In this section we define our notation and review the CP violating observables,
following the treatment of reference \cite{WP}.

   Let $M^0$ be the neutral meson (i.e. $B^0$)
and $\bar{M}^0$ its antiparticle. $M^0$
and $\bar{M^0}$ can mix with each other and form two physical mass eigenstates

\begin{equation}
M_1  =  p| M^0 > + q | \bar{M^0} >, \qquad
 M_2  =  p| M^0 > - q | \bar{M^0} >
\end{equation}
The CP-violating parameter $\epsilon_{M}$ is introduced via
\begin{equation}
\epsilon_{M} = \frac{1-q/p}{1 + q/p} \  , \qquad \frac{q}{p} \equiv \sqrt{
\frac{H_{21}}{H_{12}}}
\end{equation}
where $H_{12} \equiv M_{12} - \frac{i}{2} \Gamma_{12} =
< M^0 | H_{eff} |\bar{M}^0 > $.

  Let $f$ denote the final decay state of the neutral meson and $\bar{f}$
its charge conjugate state. The decay amplitudes of $M^0$ and $\bar{M^0}$
are denoted by
\begin{equation}
g \equiv <f|H_{eff}| M^0>, \   h  \equiv  <f|H_{eff}|\bar{M^0} >; \
\bar{g}  \equiv  <\bar{f}|H_{eff}|\bar{M^0}>, \
\bar{h}  \equiv  <\bar{f}|H_{eff}|M^0 >
\end{equation}
Parameters containing direct CP violation are defined by

\begin{equation}
\epsilon_{M}'  \equiv  \frac{1-h/g}{1+h/g}, \
 \bar{\epsilon}_{M}' \equiv \frac{1-\bar{g}/\bar{h}}{1+ \bar{g}/\bar{h}}; \
 \epsilon_{M}''  \equiv  \frac{1-\bar{g}/g}{1+ \bar{g}/g}, \   \bar{
\epsilon}_{M}'' \equiv \frac{1- h/\bar{h}}{1+ h/\bar{h}}
\end{equation}

In terms of these, the rephase-invariant observables are:

\begin{eqnarray}
a_{\epsilon} & = &  \frac{1 - |q/p|^2}{1 + |q/p|^2} =
\frac{2 Re \epsilon_{M}}{1 + |\epsilon_{M}|^2} \ ,  \qquad
a_{\epsilon'} =  \frac{1 - |h/g|^2}{1 + |h/g|^2} =
\frac{2 Re \epsilon_{M}'}{1 + |\epsilon_{M}'|^2} \ ; \nonumber \\
a_{\epsilon + \epsilon'} & = & \frac{-4 Im(qh/pg)}{(1+|q/p|^2)(1+|h/g|^2)}
= \frac{2 Im\epsilon_{M} (1-|\epsilon_{M}'|^2) + 2 Im\epsilon_{M}'
(1-|\epsilon_{M}|^2) }{ (1 + |\epsilon_{M}|^2 )(1+|\epsilon_{M}'|^2)}  \\
a_{\epsilon \epsilon'} & = & \frac{4 Re(qh/pg)}{(1+|q/p|^2)(1+|h/g|^2)} -1
= \frac{4 Im\epsilon_{M} \  Im \epsilon_{M}' - 2
(|\epsilon_{M}|^2 + |\epsilon_{M}'|^2) }{( 1 + |\epsilon_{M}|^2)
( 1 + |\epsilon_{M}'|^2)} \nonumber
\end{eqnarray}
Only three of them are independent as
 $(1- a_{\epsilon}^{2})(1 - a_{\epsilon'}^{2}) =  a_{\epsilon + \epsilon'}^{2}
 + (1+ a_{\epsilon \epsilon'})^{2}$. Analogously, one has observables
$a_{\bar{\epsilon}'}$, $a_{\epsilon + \bar{\epsilon}'}$ and
$a_{\epsilon \bar{\epsilon}'}$ similar to
$a_{\epsilon'}$, $a_{\epsilon + \epsilon'}$ and
$a_{\epsilon \epsilon'}$ but with $\epsilon_{M}'$ being replaced by
$\bar{\epsilon}_{M}'$.

  Two additional rephase-invariant quantities complete the set of
observables,

\begin{equation}
a_{\epsilon''} = \frac{1-|\bar{g}/g|^{2}}{1 + |\bar{g}/g|^2 }
= \frac{2 Re \epsilon_{M}'' }{1 + |\epsilon_{M}''|^2 }, \qquad
a_{\bar{\epsilon}''} =  \frac{1-|\bar{h}/h|^{2}}{1 +
|\bar{h}/h|^2 } = \frac{2 Re \bar{\epsilon}_{M}'' }{1 +
|\bar{\epsilon}_{M}''|^2 }
\end{equation}

 To apply this general analysis to the specific processes considered
in this work, consider the following two cases:\\

 i) \  $M^{0} \rightarrow f $ ($M^{0} \not\rightarrow \bar{f} $) , \
$\overline{M}^0 \rightarrow \bar{f}$ ($\overline{M}^0 \not \rightarrow f$)
, i.e., $f$ or $\bar{f}$ is not a common final
state of $M^{0}$ and $\overline{M}^{0}$.  This applies to the channel
 $\bar B^0 \rightarrow K^- \pi^+$.

 ii) \  $M^{0} \rightarrow (f = \bar{f}, \  f^{CP} =  f) \leftarrow
 \overline{M}^0$, i.e., final states are CP eigenstates. This applies
to the channels $\bar B^0  \rightarrow \pi^{+} \pi^{-} $, $\pi^{0} \pi^{0},
K_s \pi^0, \bar K K$.\\

In the following we will make the good approximation that $a_\epsilon=0$.
In the scenario i),    one has: $a_{\epsilon'} = - a_{\bar{\epsilon}'} = 1$
, $ a_{\epsilon + \epsilon'} = 0  = a_{\epsilon + \bar{\epsilon}'} $ and
$a_{\epsilon \epsilon'} = -1 = a_{\epsilon \bar{\epsilon}'}$. The
time-dependent rate asymmetry is:

\begin{eqnarray}
& & A_{CP}(t)  =  \frac{
\Gamma(M^0 (t) \rightarrow f) - \Gamma (\overline{M}^0 (t)\rightarrow \bar f}
{\Gamma(M^{0}(t) \rightarrow f) + \Gamma
(\overline{M}^{0} (t) \rightarrow \bar{f} )}  =
 a_{\epsilon''}
\end{eqnarray}
There is a second asymmetry corresponding to the last expression
but with $\overline{M}^0$ replaced by $M^0$.  When $a_\epsilon=0$ these
asymmetries are equal. This asymmetry also applies to charged decays.\\


In the scenario ii) in which
$a_{\epsilon'} = a_{\epsilon''}= a_{\bar{\epsilon'}}= a_{\bar{\epsilon}''}$
and  $a_{\epsilon + \epsilon'} = a_{\epsilon + \bar{\epsilon}'}$, the
time-dependent CP asymmetry is:


\begin{equation}
A_{CP}(t) \simeq
 \frac{   a_{\epsilon'}  \cos (\Delta m t) +
a_{\epsilon + \epsilon'}   \sin (\Delta m t) }{ \cosh (\Delta \Gamma t)  +
(1 + a_{\epsilon \epsilon'})  \sinh (\Delta \Gamma t)}
\end{equation}\\

\noindent If $|\Delta \Gamma |
\ll |\Delta m |$ and $|\Delta \Gamma /\Gamma | \ll 1$ then, $A_{CP}(t)$
further simplifies

\begin{equation}
A_{CP}(t) \simeq
 a_{\epsilon'}  \cos (\Delta m t) +
a_{\epsilon + \epsilon'}   \sin (\Delta m t)
\end{equation}
which may be applied,  in a good approximation, to the $B^{0} - \bar{B}^{0}$
system.\\

Because of the notation in this section, $a_\epsilon'$ is the rate asymmetry
for the CP eigenstates, case (ii) decays, whereas $a_{\epsilon''}$ is the rate
asymmetry for case (i) and charged decays.  To simplify the tables of the next
section, the rate asymmetry always appears labeled by $a_\epsilon'$ in the
tables.

\section{Results and Discussion of Rates, Asymmetries, and Sum Rules}

In Tab. 2 and 3 we present the decay parameters for the $\pi \pi$, $K \pi$
and $K K$ channels. In both tables we use the CKM parameters of ref. \cite{Ali}
with $\rho$ negative and small, $\rho=-0.12, \eta=0.34$.
  Tab. 2 is for $N = \infty$ and Tab. 3 is for
$N = 2$.  For each channel (first column) we present results for the four
cases: (t+p+e), (t+p),
(t+p') and (t), second column.  By (t) we denote tree contributions, color
dominant or color suppressed.  By (p) we denote strong penguin contributions,
including the effect of the strong phase of our model.  By (p') we denote the
penguin without the strong phase.  By (e) we denote electroweak penguin
 contributions including the absorptive parts.
One purpose of this work is to investigate the influence of the penguin
contributions with absorptive parts on the time dependent asymmetries for
the decay of neutral $B$ mesons into CP eigenstates $f = \bar f = f_{CP}$.
Such cases are the decays $\bar B^0 \rightarrow \pi^0 \pi^0, \pi^+ \pi^-,
\bar{K^0}\pi^0, (K_s\pi^0)$ and $K^0 \bar K^0$. The asymmetry $A_{CP}(t)$
(see (31)) has two terms, one proportional to $cos(\Delta m t)$ with the
coefficient $a_\epsilon'$, the other one proportional to $sin(\Delta m t)$
with coefficient $a_{\epsilon+\epsilon'}$.  These two coefficients are given
in the third and fourth columns of Tab. 2 and 3, respectively.
For the channels that are
not CP eigenstates, there is no entry because there is no time dependent
asymmetry.  In the fifth column we give the rates averaged over $B$ and
$\bar B$.
 In
the last two columns we present the real and imaginary part of the amplitude
as defined in (15).

The rates and asymmetry parameters are quite different for $N=\infty$ and
$N=2$ so we consider these cases separately, first in the approximation
of neglecting the electroweak penguins.

\subsection{$N=\infty$, Without Electroweak Penguins}
Let us consider first the results for $\bar B^0 \rightarrow \pi^0 \pi^0$
with $1/N= 0$.
Here the influence of the penguins is very large, since C (color suppressed
tree) and strong penguin P are of the same
order of magnitude and have the same sign.  The amplitude of this decay is
$\sim (P-C)$, which is small so that with strong penguins the sign of
  the real part of the
amplitude changes.  This has an effect on $a_{\epsilon+\epsilon'}$, which
 for t + p' also changes sign, so that in this case $a_{\epsilon+\epsilon'}$
changes dramatically compared to the tree value
 $a_{\epsilon+\epsilon'}=-sin 2\alpha$.
With absorptive parts added (t + p) we generate a non-zero value for
$a_\epsilon'$, which now is of the same order of magnitude as
$a_{\epsilon+\epsilon'}$ completely changing the time dependence
of $A_{CP}(t)$.  We see that $a_{\epsilon+\epsilon'}$ is influenced very
little by the absorptive part.  The branching ratios are small in all
cases, about $5\cdot 10^{-7}$ even when penguin terms are included.

The pattern is similar for the decay $\bar B^0 \rightarrow \bar K^0 \pi^0$.
The influence of the penguin terms is even stronger as one sees in the
real part of the amplitude which changes by a factor of 40, increasing the
branching ratio by two orders of magnitude.  The influence on
$a_{\epsilon+\epsilon'}$ is also strong, as in the previous case.  The other
parameter $a_\epsilon'$ is non-vanishing when we add the absorptive parts,
but is only about -3\% so that it has little influence on $A_{CP}$.

Next we discuss the decay $\bar B^0 \rightarrow \pi^+\pi^-$ which is most
interesting since it is considered as the decay channel by which
the angle $\alpha$ could be measured from the asymmetry $A_{CP}(t)$.
This is governed by the two parameters $a_\epsilon'$ and
$a_{\epsilon+\epsilon'}$.  The second parameter $a_{\epsilon+\epsilon'}=
-sin2 \alpha$ for a pure tree amplitude.  In Tab. 2 we observe that
$a_{\epsilon+\epsilon'}$ is decreased by 20\% by the strong
penguin contributions.
The absorptive parts change this additionally by only the negligible amount
of 2\% in absolute value.  The parameter $a_\epsilon'$ which determines the
$cos(\Delta m t)$ contribution of $A_{CP}(t)$ is non-negligible,
$a_\epsilon'$/$a_{\epsilon+\epsilon'}\sim -0.1$.

The last decay with a time dependent asymmetry is the pure penguin mode
$\bar B^0 \rightarrow K^0 \bar K^0$.  The parameters $a_{\epsilon+\epsilon'}$
and $a_\epsilon$ are almost equal when absorptive parts are included.  They
are of the order of 10\% so that the total time dependence is not very
large.  The change of $a_{\epsilon+\epsilon'}$  by the absorptive parts
of the penguins is only about 20\%.

In Tab.  2 we also present the results for the decay
 $\bar B^0 \rightarrow K^-\pi^+$.
 This final state is not a CP eigenstate and we must apply the
formalism of case (i) above.  The only physical parameters are the branching
ratio and the parameter $a_{\epsilon''}$ defined in (25) above.  This parameter
is listed in the third column of Tab. 2 and 3 respectively with the label
$a_{\epsilon'}$.  The asymmetry is of the order of 10\% and is negative.  As
we see, this asymmetry is time independent and not affected by mixing.  We
notice that $a_{\epsilon'}(\pi^+\pi^-) \simeq - a_{\epsilon'}(K^-\pi^+)$ to
a very good approximation.  This result can be easily explained from the
formulas and also through the recent work of Deshpande and He \cite{k7}.

The decays $B^-\rightarrow \pi^0\pi^-,K^-\pi^0, \bar K^0 \pi^-, K^0K^-$
 have been considered
in our earlier work \cite{k9}.  Of interest are the results for the branching
 ratios and the rate asymmetry parameter $a_{\epsilon'}$.  Since in this work
we assumed different values of the CKM parameters $\rho$ and $\eta$ the
asymmetry has changed.  The different sign as compared to our earlier work
has to do with the fact the the asymmetry is the difference
$B^0 - \bar B^0$ \cite{WP} rather the opposite used in ref \cite{k9}.

\subsection{N=2, Without Electroweak Penguins}
When we now look back at the more realistic case with $N=2$, which
accounts for the sign change in the QCD coefficient $a_1$ (See(18)) the
pattern of the results does not change very much, but with two exceptions.
In the decay $\bar B^0 \rightarrow \pi^0 \pi^0$ the interference between
tree and penguin contributions in the amplitude proportional to (P-C)
is now different.  This has the effect that $a_{\epsilon'}$ is smaller now and
has the opposite sign.  Furthermore the effect of the absorptive penguin
terms on $a_{\epsilon+\epsilon'}$ is decreased and the branching ratio
increases now due to penguin contributions.
In the decay $\bar B^0 \rightarrow \bar K^0 \pi^0$ the $a_\epsilon'$ is also
changed; it is smaller and has the opposite sign.  $a_{\epsilon+\epsilon'}$
is increased and the branching ratio is somewhat smaller than in the
$N=\infty$ case.
Of course the result
$a_{\epsilon'}(\pi^+\pi^-) \simeq -a_{\epsilon'}(K^-\pi^+)$ is still
valid.  In ref. \cite{k7} one can see explicitly that this
relation is independent of the details of the QCD coefficients.

We conclude that independent of details of the QCD coefficients
the influence of the penguins as compared to the tree contributions
is very significant in the channels $\pi^0\pi^0, \bar K^0 \pi^0$.
The penguins influence rates and also the time dependence through
$a_{\epsilon +\epsilon'}$.  The influence of the absorptive parts
on the time dependence is not very important. They do influence
$a_{\epsilon +\epsilon'}$ and add an additional time dependent term
in the decay $\bar B^0 \rightarrow \pi^0 \pi^0$ in the $N=\infty$
version where the C and P terms interfere destructively.

\subsection{Electroweak Penguins}

The most significant electroweak penguin operator is the term
proportional to $c_9$ which influences coefficient $a_9$ and
$a_{10}$ (much less) in (18).
Therefore we expect channels in Tab. 1 with $a_9$ coefficients
to be most strongly affected by the electroweak penguins.  This is evident
from Tab. 2 and 3 for the
$\pi^0 \pi^0, \pi^0 \pi^-, \bar K^0 \pi^0$ and $K^- \pi^0$ channels if we look
at amplitudes.  The change is stronger in $ReA$.  The  change of the imaginary
part is only 30\% of the change of the real part for the $\pi^0 \pi^0$ and
$\pi^0 \pi^-$ and negligible in the imaginary part for the $\bar K^0 \pi^0$ and
$K^- \pi^0$ channels. This is because the CKM phase is real in the $K\pi$
channels.  It is interesting to note that
for $N=\infty$ the time dependence of the asymmetry for the
decay $\bar B^0 \rightarrow \pi^0 \pi^0$ is significantly influenced by the
 change of
$a_{\epsilon+\epsilon'}$.  This is not the case for $N=2$, where
$a_{\epsilon'}$
is also smaller.  From Tab. 2 and 3 it is evident that the electromagnetic
penguins can influence branching ratios substantially as shown by the
$K^-\pi^0$ and $\bar K^0 \pi^0$ channels.

\subsection{Effect of Penguins on Flavor Sum Rules}

In this section we shall examine the effects of the SU(3) breaking in the
matrix elements and the influence of the EW penguins on the sum rules of
\cite{GL,NQ,GRL}.  We shall start with the sum rule for $B\rightarrow
\pi\pi$ amplitudes based on SU(2) symmetry \cite{GL},  which with our
definition
of quark content in (19-21) is:

\be A=B \ee

\noindent where

\be
A=\sqrt{2}A(B^-\rightarrow \pi^-\pi^0)\ee

\be
B=\sqrt{2} A(\bar B^0\rightarrow \pi^0\pi^0) - A(\bar B^0 \rightarrow
 \pi^+\pi^-)\ee

If we neglect the $\pi^\pm-\pi^0$ mass differences etc, the amplitude
$A$ and $B$ have according to Tab. 1 the following form in terms of the
coefficients $\hat a_i$ which contain in addition to the Wilson coefficients
$a_i$ the appropriate CKM factors for tree and penguin contributions,
\be
A=B=M\{-(\hat a_1+ \hat a_2) +
\frac{3}{2}(\hat a_7 - \hat a_9)-\frac{3}{2}(\hat a_{10} +
  \hat a_8R[\pi\pi])\}
\ee
where $M$ is the reduced matrix element assuming SU(2) symmetry and
factorization which can be read off from (16) and Tab. 1.  This means that
the sum rule $A=B$ is not spoiled by penguin contributions even when the EW
penguins are included.  In (34) $\hat a_2(\hat a_1)$ can be identified with
 the tree
(color suppressed tree) and
$\frac{3}{2}(\hat a_7-\hat a_9)[\frac{3}{2}(\hat a_{10}+\hat a_{8}R)$]
with the EW penguin [color suppressed EW penguin].  That $A=B$ remains valid
with
the EW penguin terms stems from the  fact that the additional contributions
proportional to $O_{7...10}$ transform as the same mixture of I=1/2 and 3/2
operators as the tree operators $O_{1,2}$. The strong penguins on the other
hand have only a single isospin component. This explains why the amplitudes
$A$ and $B$ have no strong penguin contributions.  In Tab. 4 and 5 we give
the numerical results of the amplitudes $A$ and $B$ in our model. As we see
 $A=B$
is very well satisfied in all four versions of the model calculation: t, t+p',
t + p and t+p+e.  By comparing the results for t+p+e with t+p it is clear that
the electroweak penguins make substantial contributions ($\sim 15\%$ in $ReA,
\sim 2\%$ in $ImA)$ but preserve the sum rule for the reason given above.
There are therefore substantial I=3/2 pieces in the amplitudes $A$ and $B$
which do not originate from the tree operators $O_{1,2}$.

Similarly, we have for the $B\rightarrow K\pi$ transitions the sum rule
\be C=D \ee
where
\be C=[\sqrt{2}A(\bar B^0\rightarrow \bar K^0\pi^0) -
A(\bar B^0\rightarrow  K^-\pi^+)]/r_u\ee

\be D=[\sqrt{2}A( B^-\rightarrow  K^-\pi^0) +
A( B^-\rightarrow  \bar K^0\pi^-)]/r_u\ee

\noindent where $r_u=V_{us}/V_{ud}$ has been inserted for later use.

Since for $B\rightarrow K\pi$  two different reduced matrix elements occur
in (16), the amplitudes $C$ and $D$ are now given by
\be
C=D=M_2/r_u\{-\hat a'_1+\frac{3}{2}(\hat a'_7-\hat a'_9)\} + M_1/r_u\{-\hat
a'_2
-\frac{3}{2}(\hat a'_{10} + \hat a'_8 R[K,\pi])\}\ee
where the $M_1(M_2)$ are the reduced matrix elements in (16) proportional to
the $Z_1^{(q)}$($Z_2^{(q)}$) and the $\hat a'_i$ are the Wilson coefficients
multiplied by the appropriate CKM matrix elements.  As in the case $A=B$  the
sum rule $C=D$ is broken only by small SU(2) mass differences. Futhermore the
strong penguin amplitudes drop out in the amplitudes $C$ and $D$ for the same
reason they did in $A$ and $B$.  In Tab. 4 and 5 we can see that $C=D$ is quite
well satisfied in all four model versions except perhaps for the $ReA$ in t+p
which is caused by the symmetry breaking effects in connection with
$O(\alpha_s)$ corrections.  By comparing the values for t+p+e with t+p we
observe that the EW penguin effects are very large in $ReA$ (factor $\sim 3$)
but again preserve the sum rule $C=D$.

An important sum rule based on SU(3) symmetry and absence of EW penguins
in reference \cite{GRL} is
\be A=C \ee
In contrast to the previously considered sum rules, different matrix elements
are now involved, primarily those with different factors $f_K/f_\pi \ne 1$.  If
we compare (34) with (38) it is clear that the sum rule $A=C$ is valid
if $M=M_1/r_u=M_2/r_u$ and if there are no EW penguins. Thus when EW penguins
are neglected, we would expect $(f_K/f_\pi) A = C$ which is approximately valid
in Tab. 4 and 5.  The factor $1/r_u$ corrects for the different CKM factors in
the $b\rightarrow d$ versus $b\rightarrow s$ transitions in the tree
amplitudes.  Because the electroweak penguin terms in (39) have a different
CKM factor this sum rule is strongly violated by the EW penguin.  Indeed if we
compare $(f_K/f_\pi)A$ with $C$ for t+p+e we observe a factor of 3.5
difference in $ReA$.

The last relation considered in reference \cite{GRL} is the quadrangle
relation $E=F$.  For our model including strong and EW penguins we have

\be
E=(\hat a_2 + \hat a_4 + \hat a_6 R[\pi,\pi] +\hat a_{10} +\hat a_8R[\pi,\pi])M
-( \hat a_4 + \hat a_6 R[K,K]  - \frac{1}{2}(\hat a_{10} +\hat a_8R[K,K]))M_K
\ee

\noindent and

\be
F=[\hat a'_2 +\frac{1}{2}(\hat a'_{10} + \hat a'_8 R[K,\pi])]M_1/r_u\ee

It is clear from (40,41) that the relation $E=F$ is badly spoiled by the SU(3)
symmetry breaking in the reduced matrix elements, i.e. $M \ne M_K \ne M_1$ and
$R[\pi,\pi] \ne R[K,K] \ne R[K,\pi]$.  The EW penguins in the amplitude
$E$ are very small since
they are color suppressed and Cabibbo suppressed compared to the tree term
(compare line t+p+e with t+p in Tab. 4 and 5 for case E).  In contrast the
EW penguin of the amplitude $F$ while color suppressed is not Cabibbo
suppressed,
which explains the larger difference by almost a factor of two.
  This has the consequence that apart from the SU(3) breaking the sum rule
$E=F$ is very badly violated for the $ReA$ by a factor of 2.5 ($N=\infty$) or
1.7 ($N=2$), respectively.

This calculation supports the criticism of reference \cite {k6,k7} to the
scheme for extracting weak phases advocated in reference \cite{GRL}.  The
latter
authors have subsequently modified their analysis to include EW penguins
\cite{GRLpen}.  By making additional measurements of $B_s$ decays,
 they argue that it is still possible  to extract the CKM angles
from rate measurements alone.   A somewhat different scheme has been proposed
by Deshpande and He \cite{Desheta} involving the additional measurement of
the rate for $B^- \rightarrow K^- \eta_8$ and the assumption of no EW penguin
contributions in $B^- \rightarrow \pi^0 \pi^-$;  as we see in our model this
is valid up to
$15\%$ for $ReA$ and $2\%$ in $ImA$, as shown in Tab. 4 and 5.

\section{Summary and Conclusions}

We have reported the results of a model of trees, strong and EW penguins with
absorptive corrections.  This model is subject to various well known
uncertainties associated with soft physics.  In addition we had to use
badly determined CKM parameters.  Nevertheless we believe our results give
the right order of magnitude of the effect of strong and EW penguins and
absorptive corrections.  They indicate that certain flavor sum rules are
spoiled by the EW penguins and SU(3) symmetry violation.  As expected, SU(2)
sum rules are preserved, even those with EW penguin  effects.  Strong and weak
penguins have important effects on the branching ratios and the CP violation
parameters, with details depending on the model of the effective Wilson
coefficients ($N=\infty$ or $N= 2$).  By calculating more rates one could
test the procedures of Deshpande and He \cite{k6,Desheta} and Gronau et. al.
\cite{GRLpen} for determining the CP violating phase $\gamma$.

\subsection*{Acknowledgement}
W.\,F.\,P. thanks the Desy Theory Group for its kind hospitality and the North
Atlantic Treaty Organization for a Travel Grant.


\newpage

\section*{Table captions}
\begin{description}

\item Tab. 1a,b:  Factorization coefficients $Z^{(q)}_{1,2}$ for various
$B\to P_1 P_2$ decays. The short distance coefficients $a_i$ are defined
in \eq(\ref{def_of_ai}) and the factor $R$ is given in \eq(\ref{def_of_R}).
The coefficients do not include the appropriate CKM phases as in (1) and (2).

\item Tab. 2: Decay and CP violating parameters
 in $B \to \pi\pi, \pi K, K \bar{K}$.  The amplitudes are calculated in a model
with tree (t) and strong penguins p (p') with (without) absorptive parts
and electroweak penguins (e) with absorptive parts. The model parameters are
 NLL QCD Coefficients, BSW model for matrix elements,
Wolfenstein parameters: $(\rho,\eta)=(-0.12, 0.34)$, $N = \infty $.

\item Tab. 3: Decay and CP violating parameters
 in $B \to \pi\pi, \pi K, K \bar{K}$.  The amplitudes are calculated in a model
with tree (t) and strong penguins p (p') with (without) absorptive parts
and electroweak penguins (e) with absorptive parts. The model parameters are
 NLL QCD Coefficients, BSW model for matrix elements,
Wolfenstein parameters: $(\rho,\eta)=(-0.12, 0.34)$, $N = 2 $.

\item Tab. 4: Sum rules for reduced
$B \to \pi\pi, \pi K, K \bar{K}$ amplitudes. The amplitudes are calculated in a
model with tree (t) and strong penguins p (p') with (without) absorptive parts
and electroweak penguins (e) with absorptive parts. The model parameters are
NLL QCD Coefficients, BSW model for matrix elements, Wolfenstein parameters:
$(\rho,\eta)=(-0.12, 0.34)$, $N = \infty $.

\item Tab. 5:  Sum rules for reduced
$B \to \pi\pi, \pi K, K \bar{K}$ amplitudes. The amplitudes are calculated in a
model with tree (t) and strong penguins p (p') with (without) absorptive parts
and electroweak penguins (e) with absorptive parts. The model parameters are
NLL QCD Coefficients, BSW model for matrix elements, Wolfenstein parameters:
$(\rho,\eta)=(-0.12, 0.34)$, $N = 2$.

\end{description}

\newpage

\def\tmtwo{$\times10^{-2}$}
\def\tmthree{$\times10^{-3}$}
\def\tmfour{$\times10^{-4}$}
\def\tmfive{$\times10^{-5}$}
\def\tmsix{$\times10^{-6}$}
\def\tmseven{$\times10^{-7}$}
\def\tmeight{$\times10^{-8}$}

\def\nc{\\}
\def\sc{\\[-1mm]}
\centerline{Tab. 1a}

\begin{center}
\begin{tabular}{||c|c|c||}
\hline \hline
$P_1$ & $P_2$ & $Z^{(q)}_1$  \\ \hline
$\bar{K}^0$&$\pi^0$    & $(a_4 -a_{10}/2+( a_6-a_8/2) \,R[\bar {K}^0,\pi^0])/
\sqrt{2}$
                    \\
$K^-$ & $\pi^+$     & $a_2\delta_{qu} + a_4 + a_{10} +(a_6+a_8)
 \,R[K^-,\pi^+]$
                    \\
$K^-$ & $\pi^0$     & $-(a_2 \d_{qu} + a_4 +a_{10}+ (a_6+a_8) \,R[K^-,\pi^0])/
\sqrt{2}$
                     \\
$\bar{K}^0$&$\pi^-$ & $a_4 -a_{10}/2+ (a_6-a_8/2) \,R[K^0,\pi^-]$
                     \\
$\pi_1^0$ & $\pi_2^0$& $(-a_1 \d_{qu} + a_4 -a_{10}/2+3a_7/2-3a_9/2+
 (a_6-a_8/2) \,R[\pi_1^0,\pi_2^0])/2$
                      \\
$\pi^+$ & $\pi^-$   & $0$
                     \\
$\pi^-$ & $\pi^0$   & $(-a_2 \d_{qu} - a_4 -a_{10}- (a_6+a_8) \,
R[\pi^-,\pi^0])/\sqrt{2}$
                      \\
$K^0$ & $K^-$       & $a_4 -a_{10}/2+ (a_6-a_8/2) \,R[K^0,K^-]$
                     \\
$K^0$ & $\bar K^0$       & $a_4 -a_{10}/2+ (a_6-a_8/2) \,R[K^0,\bar K^0]$
                     \\
\hline
\end{tabular}
\end{center}

\centerline{Tab. 1b}

\begin{center}
\begin{tabular}{||c|c|c||}
\hline \hline
$P_1$ & $P_2$ &  $Z^{(q)}_2$ \\ \hline
$\bar{K}^0$&$\pi^0$
                    & $(-a_1\d_{qu}+3a_7/2-3a_9/2)\sqrt{2}$\\
$K^-$ & $\pi^+$
                    & $0$\\
$K^-$ & $\pi^0$
                    & $(-a_1\d_{qu}+3a_7/2-3a_9/2)  /\sqrt{2}$ \\
$\bar{K}^0$&$\pi^-$
                    & $0$ \\
$\pi_1^0$ & $\pi_2^0$
                    & $(-a_1 \d_{qu}  +a_4 -a_{10}/2+3a_7/2-3a_9/2+
 (a_6-a_8/2) \,
R[\pi_2^0,\pi_1^0])/2$  \\
$\pi^+$ & $\pi^-$
                    & $a_2 \d_{qu} +a_4 +a_{10}+ (a_6+a_8) \,
R[\pi^-,\pi^+]$  \\
$\pi^-$ & $\pi^0$
                    & $(-a_1 \d_{qu} + a_4 +3a_7/2-3a_9/2-a_{10}/2+
 (a_6-a_8/2) \,R[\pi^0,\pi^-])/\sqrt{2}$  \\
$K^0$ & $K^-$
                    & $0$ \\
$K^0$ & $\bar K^0$
                    & $0$ \\
\hline
\end{tabular}
\end{center}

\renewcommand{\baselinestretch}{1.1}

\eject
\centerline{Tab. 2}
\def\tmtwo{$\times10^{-2}$}
\def\tmthree{$\times10^{-3}$}
\def\tmfour{$\times10^{-4}$}
\def\tmfive{$\times10^{-5}$}
\def\tmsix{$\times10^{-6}$}
\def\tmseven{$\times10^{-7}$}
\def\tmeight{$\times10^{-8}$}

\footnotesize
\begin{center}
\begin{tabular}{||l||l||l|l|l|l|l||}
\hline\hline
\multicolumn{7}{||c||}{{Decay Parameters and CP Violation in $B \to \pi\pi, \pi
K, K \bar{K}$}}\\
\multicolumn{7}{||c||}{Tree, Strong and EW Penguins t,p (p'), e with (without)
    Absorptive Parts} \\
\multicolumn{7}{||c||}{{$N = \infty $ NLL QCD Coefficients, BSW model,
                               $(\rho, eta)=(-0.12, 0.34)$}}\\
\hline
           &  &\multicolumn{5}{|c||}{\bf{Decay Parameters}}\\
\hline
Channel
&Amp&$a_{\epsilon'}$&$a_{\epsilon+\epsilon'}$&$<BR>$
                                           &Re~A$\times 10^3$&Im~A$\times
 10^3$\\
\hline \hline
          &t + p + e&0.308 &-0.126&4.22\tmseven   &0.256&-1.40    \\
            &t + p &0.316 &0.201 &4.25\tmseven   &0.528&-1.32\\
$\pi^0\pi^0$&t + p'&0.0   &0.216 &4.14\tmseven   &0.654&-1.56\\
           &t      & 0.0  &-0.951&5.66\tmseven  &-0.658&-1.86 \\
\hline
          &t + p + e&0.0703 &-0.784 &1.19\tmfive   &-1.65 &-8.59    \\
           &t + p  &0.0708&-0.766 &1.19\tmfive   &-1.59 &-8.57\\
$\pi^+\pi^-$&t + p'&0.0   &-0.777 &1.18\tmfive   &-1.41&-8.93\\
           &t      &0.0   &-0.951 &1.43\tmfive   &-3.31  &-9.37\\
\hline
          &t + p + e&0.0000&--   &3.50\tmsix     &1.46  &4.69    \\
           &t + p  &0.0   &--     &3.69\tmsix    &1.68  &4.76\\
$\pi^0\pi^-$&t + p'&0.0   &--     &3.69\tmsix    &1.68  &4.76\\
           &t      &0.0   &--     &3.69\tmsix    &1.68  &4.76 \\
\hline
          &t + p + e&-0.0365&0.435 &5.49\tmsix  &-6.09 &-1.56    \\
           &t + p&-0.0265 &0.457  &7.78\tmsix  &-7.29 &-1.54\\
$\bar{K}^0\pi^0$&t + p'&0.0&0.454 &7.58\tmsix  &-7.25&-0.384\\
           &t    &0.0     &-0.951 &3.68\tmeight&-0.168&-0.476\\
\hline
          &t + p +e&-0.0760&--     &1.74\tmfive  &-10.6   &-4.11    \\
           &t + p&-0.0725 &--     &1.82\tmfive &-10.9   &-4.10\\
${K^-}\pi^+$&t + p'&0.0   &--     &1.79\tmfive &-10.8 &-2.47\\
           &T   &0.0      &--      &9.28\tmseven&-0.909  &-2.60\\
\hline
          &t + p + e&-0.0458&--    &11.1\tmsix  &8.66   &2.41    \\
           &t + p&-0.0587&--     &8.56\tmsix  &7.56    &2.43\\
${K^-}\pi^0$&t + p'&0.0   &--     &8.37\tmsix  &7.52   &1.27\\
           &t        &0.0 &--     &3.01\tmseven&0.481   &1.36 \\
\hline
\multicolumn{7}{||c||}{{Pure Penguin Modes}}\\
\hline
          &   p + e&0.0922 &0.0984  &1.27\tmsix &2.58   &1.18 \\
           & p    &0.0937&0.0951  &1.23\tmsix &2.54   &1.17\\
$K^0\bar{K^0}$& p' &0.0   &0.118  &1.18\tmsix &2.81   &0.654\\
\hline
          &   p + e&0.0920  &--    &1.30\tmsix       &2.62   &1.20    \\
           &  p   &0.0936  &--    &1.27\tmsix  &2.58   &1.19\\
$K^0{K^-}$ & p'   &0.0     &--    &1.22\tmsix  &2.85  &0.664\\
\hline
          &   p + e&-0.00606 &--   &1.52\tmfive &-10.2  &-1.51    \\
           & p    &-0.00616 &--   &1.48\tmfive &-10.1 &-1.51\\
$\bar{K^0}\pi^-$& p'&0.0    &--   &1.44\tmfive &-10.0 &0.131\\
\hline \hline
\end{tabular}
\end{center}
\eject

\centerline{Tab. 3}
\def\tmtwo{$\times10^{-2}$}
\def\tmthree{$\times10^{-3}$}
\def\tmfour{$\times10^{-4}$}
\def\tmfive{$\times10^{-5}$}
\def\tmsix{$\times10^{-6}$}
\def\tmseven{$\times10^{-7}$}
\def\tmeight{$\times10^{-8}$}
\begin{center}
\begin{tabular}{||l||l||l|l|l|l|l||}
\hline\hline
\multicolumn{7}{||c||}{{Decay Parameters and CP Violation in $B \to \pi\pi, \pi
K, K \bar{K}$}}\\
\multicolumn{7}{||c||}{Tree, Strong and EW Penguins t,p (p'), e with (without)
    Absorptive Parts} \\
\multicolumn{7}{||c||}{{$N=2$ NLL QCD Coefficients, BSW model,
                               $(\rho, eta)=(-0.12, 0.34)$}}\\
\hline
           &  &\multicolumn{5}{|c||}{\bf{Decay Parameters}}\\
\hline
Channel
&Amp&$a_{\epsilon'}$&$a_{\epsilon+\epsilon'}$&$<BR>$
                                           &Re~A$\times 10^3$&Im~A$\times
 10^3$\\
\hline \hline
            &t + p + e&-0.0834&-0.938 &6.11\tmseven &1.17    &1.79 \\
            &t + p   &-0.0615 &-0.877 &7.76\tmseven &1.47    &1.88\\
$\pi^0\pi^0$&t + p'  &0.0     &-0.881 &7.69\tmseven &1.56    &1.70 \\
           &t        &0.0     &-0.951 &3.41\tmseven &0.511   &1.45     \\
\hline
            &t + p + e&0.0620  &-0.781 &8.75\tmsix  &-1.37   &-7.40 \\
           &t + p    &0.0609  &-0.793 &8.83\tmsix  &-1.46    &-7.43 \\
$\pi^+\pi^-$&t + p'  &0.0     &-0.794 &8.82\tmsix  &-1.32   &-7.69 \\
           &t        &0.0     &-0.951 &1.06\tmfive &-2.84   &-8.05\\
\hline
            &t + p + e&0.000399&--     &7.83\tmsix &2.14    &7.03 \\
           &t + p    &0.0     &--     &8.31\tmsix &2.52    &7.14\\
$\pi^0\pi^-$&t + p'  &0.0     &--     &8.31\tmsix &2.52    &7.14\\
           &t        &0.0     &--     &8.31\tmsix &2.52    &7.14\\
\hline
            &t + p + e&0.0264 &0.697  &2.70\tmsix   &-4.26    &-0.444 \\
           &t + p    &0.0145 &0.670  &4.58\tmsix  &-5.59    &-0.424 \\
$\bar{K}^0\pi^0$&t + p'&0.0   &0.672  &4.48\tmsix  &-5.56      &0.446 \\
           &t        &0.0     &-0.951 &2.21\tmeight&0.131      &0.370\\
\hline
            &t + p + e&-0.0659&--     &1.30\tmfive  &-9.22    &-3.35 \\
           &t + p    &-0.0709&--     &1.20\tmfive   &-8.83     &-3.35 \\
${K^-}\pi^+$&t + p'  &0.0     &--     &1.18\tmfive   &-8.79     &-2.13  \\
           &t        &0.0     &--     &8.09\tmseven  &-0.789    &-2.24     \\
\hline
            &t + p + e&-0.0560 &--     &9.39\tmsix   &7.85     &2.72 \\
           &t + p    &-0.0809  &--     &6.40\tmsix   &6.38     &2.74  \\
${K^-}\pi^0$&t + p'  &0.0     &--     &6.29\tmsix   &6.35     &1.88         \\
           &t        &0.0     &--     &6.16\tmseven&0.689     &1.95        \\
\hline
\multicolumn{7}{||c||}{{Pure Penguin Modes}}\\
\hline
            &p + e&0.0886     &0.0959&7.57\tmseven    &2.00      &0.907 \\
           & p    &0.0862     &0.0940&7.95\tmseven  &2.06      &0.925   \\
$K^0\bar{K^0}$& p' &0.0       &0.110  &7.68\tmseven &2.26      &0.536 \\
\hline
            &p + e&0.0883     &--     &7.82\tmseven &2.04      &0.922 \\
           &  p   &0.0860     &--     &8.21\tmseven &2.09      &0.939\\
$K^0{K^-}$ & p'   &0.0        &--     &7.94\tmseven &2.30      &0.546\\
\hline
            &p + e&-0.00584   &--       &9.09\tmsix  &-7.90     &-1.12 \\
           & p    &-0.00569    &--     &9.54\tmsix   &-8.09      &-1.12\\
$\bar{K^0}\pi^-$& p'&0.0      &--     &9.32\tmsix   &-8.05      &0.108\\
\hline \hline
\end{tabular}
\end{center}
\eject
\centerline{Tab. 4}
\begin{center}
\begin{tabular}{||l|l|l|l||}
\hline\hline
\multicolumn{4}{||c||}{{Sum Rules for Reduced $B \to \pi\pi, \pi K, K \bar{K}$
 Amplitudes}}\\
\multicolumn{4}{||c||}{Tree, Strong and EW Penguins t,p (p'), e with (without)
    Absorptive Parts} \\
\multicolumn{4}{||c||}{{$N=\infty$ NLL QCD Coefficients, BSW model,
                               $(\rho, \eta)=(-0.12, 0.34)$}}\\
\hline
Channel                    &Amp           &ReA$\times 10^3$ &ImA$\times
10^3$~\\
\hline \hline
\multicolumn{4}{||c||}{{Sum Rules: A=B=C=D and E=F}}\\
\hline
\hline
                           &t + p + e      &2.06          &6.63 \\
                           &t + p          &2.37          &6.73\\
$A=\sqrt{2}\pi^0\pi^-$       &t + p'         &2.37           &6.73\\
                           &t              &2.37           &6.73 \\
\hline
                           &t + p + e      &2.01          &6.61 \\
      &t + p         &2.34           &6.70\\
$B=\sqrt{2}\pi^0\pi^0 - \pi^+\pi^-$&t+ p'   &2.34           &6.72\\
                           &t             &2.37           &6.74\\
\hline
                           &t + p + e      &8.78          &8.40 \\
                            &t + p          &3.20     &8.49\\
$C=(\sqrt{2}\bar{K}^0\pi^0 - {K^-}\pi^+)/r_u$
                            &t + p'         &2.70      &8.50\\
                            &t              &2.96      &8.53\\
\hline
                           &t + p + e      &8.95              &8.40 \\
                            &t + p          &2.80     &8.54\\
$D=(\sqrt{2}K^-\pi^0 + \bar{K^0} \pi^-)/r_u$
                            &t + p'         &2.60      &8.54\\
                            &t              &3.00      &8.44\\
\hline
\hline
                           &t + p + e      &-4.27              &-9.79 \\
                            &t + p          &-4.17     &-9.76\\
$E=\pi^+\pi^- - K^-K^0$        &t + p'         &-4.26      &-9.59\\
                            &t              &-3.31      &-9.37\\
\hline
                           &t + p + e      &-1.70              &-11.45 \\
                            &t + p          &-3.70     &-11.44\\
$F=(K^-\pi^+ - \bar{K^0}\pi^-)/r_u$&t + p'     &-3.40      &-11.47\\
                            &t              &-4.01      &-11.5\\
\hline
\end{tabular}
\end{center}
\eject
\centerline{Tab. 5}
\begin{center}
\begin{tabular}{||l|l|l|l||}
\hline\hline
\multicolumn{4}{||c||}{{Sum Rules for Reduced $B \to \pi\pi, \pi K, K \bar{K}$
 Amplitudes}}\\
\multicolumn{4}{||c||}{Tree, Strong and EW Penguins t,p (p'), e with (without)
    Absorptive Parts} \\
\multicolumn{4}{||c||}{{$N=2$ NLL QCD Coefficients, BSW model,
                               $(\rho,\eta)=(-0.12, 0.34)$}}\\
\hline
Channel                    &Amp           &ReA$\times 10^3$ &ImA$\times
10^3$~\\
\hline \hline
\multicolumn{4}{||c||}{{Sum Rules: A=B=C=D and E=F}}\\
\hline
\hline
                           &t + p + e      &3.03          &9.94 \\
                           &t + p          &3.56          &10.10\\
$A=\sqrt{2}\pi^0\pi^-$       &t + p'       &3.56          &10.10\\
                           &t              &3.56          &10.10 \\
\hline
                           &t + p + e      &3.02          &9.93 \\
      &t + p          &3.54          &10.09\\
$B=\sqrt{2}\pi^0\pi^0 - \pi^+\pi^-$&t+ p'  &3.53          &10.09\\
                           &t              &3.57          &10.10\\
\hline
                           &t + p + e      &14.10          &12.01 \\
                            &t + p         &4.08           &12.13\\
$C=(\sqrt{2}\bar{K}^0\pi^0 - {K^-}\pi^+)/r_u$
                            &t + p'        &4.09           &12.18\\
                            &t             &4.30           &12.20\\
\hline
                           &t + p + e      &14.13          &12.03 \\
                            &t + p         &4.12           &12.16\\
$D=(\sqrt{2}K^-\pi^0 + \bar{K^0}\pi^-)/r_u$
                            &t + p'        &4.10           &12.20\\
                            &t             &4.30           &12.17       \\
\hline
\hline
                           &t + p + e      &-3.41          &-8.32 \\
                            &t + p         &-3.55          &-8.37\\
$E=\pi^+\pi^- - K^-K^0$        &t + p'     &-3.62          &-8.24\\
                            &t             &-2.84          &-8.05\\
\hline
                           &t + p + e      &-5.83          &-9.84 \\
                            &t + p         &-3.27          &-9.84\\
$F=(K^-\pi^+ - \bar{K^0}\pi^-)/r_u$&t + p' &-3.26          &-9.88\\
                            &t             &-3.48          &-9.89\\
\hline
\end{tabular}
\end{center}
\eject

\end{document}